\documentclass[]{aastex631}
\usepackage{hyperref}

\usepackage[utf8]{inputenc}

\begin{document}

\shorttitle{ Fine Structure Loops}
\shortauthors{Bura et al.}

\title{Unveiling the Dynamics and Genesis of Small-scale Fine Structure Loops in the Lower Solar Atmosphere}

\correspondingauthor{Tanmoy Samanta}
\email{tanmoy.samanta@iiap.res.in}

\author[0009-0000-5018-9735]{Annu Bura}
\affiliation{Indian Institute of Astrophysics, Koramangala, Bangalore 560034, India}

\author[0000-0002-9667-6392]{Tanmoy Samanta}
\affiliation{Indian Institute of Astrophysics, Koramangala, Bangalore 560034, India}

\author{Alphonse Sterling}
\affiliation{NASA Marshall Space Flight Center, Huntsville, AL 35812, USA}

\author{Yajie Chen}
\affiliation{Max-Planck Institute for Solar System Research, 37077 Gottingen, Germany}

\author{Jayant Joshi}
\affiliation{Indian Institute of Astrophysics, Koramangala, Bangalore 560034, India}

\author{Vasyl Yurchyshyn}
\affiliation{Big Bear Solar Observatory, New Jersey Institute of Technology, 40386 North Shore Lane, Big Bear, CA 92314, USA}

\author{Ronald L. Moore}
\affiliation{NASA Marshall Space Flight Center, Huntsville, AL 35812, USA}
\affiliation{Center for Space Plasma and Aeronomic Research, University of Alabama in Huntsville, Huntsville, AL 35805, USA}

\begin{abstract}
Recent high-resolution solar observations have unveiled the presence of small-scale loop-like structures in the lower solar atmosphere, often referred to as unresolved fine structures, low-lying loops, and miniature hot loops. These structures undergo rapid changes within minutes, and their formation mechanism has remained elusive. In this study, we conducted a comprehensive analysis of two small loops utilizing data from the Interface Region Imaging Spectrograph (IRIS), the Goode Solar Telescope (GST) at Big Bear Solar Observatory, and the Atmospheric Imaging Assembly (AIA) and the Helioseismic Magnetic Imager (HMI) onboard the Solar Dynamics Observatory (SDO), aiming to elucidate the underlying process behind their formation. The GST observations revealed that these loops, with lengths of $\sim$3.5~Mm and heights of $\sim$1~Mm, manifest as bright emission structures in H$\alpha$ wing images, particularly prominent in the red wing. IRIS observations showcased these loops in 1330~\AA\ slit-jaw images, with TR and chromospheric line spectra exhibiting significant enhancement and broadening above the loops, indicative of plasmoid-mediated reconnection during their formation. Additionally, we observed upward-erupting jets above these loops across various passbands. Furthermore, differential emission measurement analysis reveals an enhanced emission measure at the location of these loops, suggesting the presence of plasma exceeding 1~MK. Based on our observations, we propose that these loops and associated jets align with the minifilament eruption model. Our findings suggest a unified mechanism governing the formation of small-scale loops and jets akin to larger-scale X-ray jets.
\end{abstract}


\section{Introduction} \label{sec:intro}
Active regions (ARs) on the Sun, characterized by strong magnetic fields, are the primary origin locations of various solar activities, including flares, coronal mass ejections (CMEs), loops, brightenings and active-region jets. Coronal loops are large structures in the AR solar corona which exhibit a wide length range, spanning from a few million meters to a substantial fraction of the solar radius. These loops likely form from the emergence of undulatory flux tubes from below the photosphere, with magnetic reconnection playing a crucial role in their formation  (\citealt{pariat2004resistive}, \citealt{he2010reconfiguration}, \citealt{tripathi2021transient}, \citealt{hou2021formation}). Magnetic reconnection occurs throughout the solar atmosphere and results in various solar activities (\citealt{dere1991explosive}, \citealt{innes1997bi}, \citealt{harra2003evidence}, \citealt{shibata2007chromospheric}, \citealt{hara2011plasma}, \citealt{tian2014prevalence}, \citealt{li2016magnetic}, \citealt{tian2018magnetic}, \citealt{hou2021formation}, \citealt{cheng2023ultra}), including formation of some loops and the generation of jets (\citealt{shibata1993observations}, \citealt{panesar2016magnetic}, \citealt{sterling2016minifilament}). Therefore, understanding the physical context in which reconnection occurs is essential for comprehending the formation of small-scale features such as loops and jets.\\
Recently, small-scale loops varying from sub-million meters to a few million meters in length have been observed in the transition region (TR) of the Sun (\citealt{hansteen2014unresolved}, \citealt{2015ApJ...810...46H}, \citealt{brooks2016properties}, \citealt{pereira2018chromospheric}), suggesting that they are a key component of small-scale solar phenomena. Similar small-scale structures have previously been referred to as ``Miniature Coronal Loops'' (\citealt{peter2013structure}), ``Unresolved Fine Structures (UFS)'' (\citealt{hansteen2014unresolved}), ``Low-Lying Loops'' (\citealt{pereira2018chromospheric}), ``Campfires'' (\citealt{berghmans2021extreme}) or ``Loop-Like Structures'' (\citealt{skan2023small}), depending on the observables in which they are detected. These structures are very dynamic and bright with respect to their background (\citealt{hansteen2014unresolved}) and have a lifetime of a few minutes (\citealt{hansteen2014unresolved}, \citealt{brooks2016properties}). Although these small-scale loops are found heated to different temperature ranges, their exact formation mechanism remains unclear.  \\
The presence of small-scale loops in the TR that do not have coronal counterparts was proposed initially by \cite{feldman1983unresolved} based on Skylab data. Using High-resolution Coronal Imager (Hi-C, \citealt{kobayashi2014high}) observations, \cite{peter2013structure} found miniature loops reaching 1.5~MK coronal temperatures. These features are $\sim$1~Mm in length and 0.2~Mm in thickness, spanning just a single granule. A subsequent study by \cite{barczynski2017miniature} examined three scenarios for miniature loop-like structures in the AR and found that these are tiny versions of hot coronal loops that can reach coronal temperatures during strong heating. The advent of the Interface Region Imaging Spectrograph (IRIS, \citealt{de2014interface}) offered an opportunity to observe the TR with high spatial and temporal resolution. \cite{hansteen2014unresolved} utilized IRIS to identify dynamic low-lying loops in the TR, which were long-postulated UFS. These low-lying loops, with lengths of 2–6~Mm and heights between 1–4.5~Mm above the solar surface, exhibit rapid velocity shifts ($>$ 80~km s$^{-1}$). In a study by \cite{brooks2016properties}, the authors examined the properties of a sample of over a hundred loops and found that the observed spatial scales, lifetimes, and heating patterns of these small features align well with the scenario of single-strand heating. \cite{pereira2018chromospheric} discovered that low-lying loops in the quiet-Sun at TR temperatures exhibit chromospheric counterparts, which are distinguished by strong Doppler shifts rather than intensity enhancements. These loops are clearly observed as absorption features appearing in the far blue or red wings of H$\alpha$. The authors suggested that magnetic reconnection might be the driver of these low-lying loops, but they did not find any signs of heating near the footpoints in chromospheric observations. The Extreme Ultraviolet Imager (EUI, \citealt{rochus2020solar}) on board the Solar Orbiter (\citealt{muller2020solar}) observed a localised brightening in the 174~\AA\ passband with length scales between 0.4-4~Mm and height between 1-5~Mm from the photosphere. These features, named ``Campfires", are mostly observed in coronal temperatures (1-1.6~MK). \cite{berghmans2021extreme} proposed that campfires are the apexes of low-lying small-scale network loops in the quiet-Sun atmosphere. Using the Helioseismic and Magnetic Imager (HMI, \citealt{scherrer2012helioseismic}) onboard Solar Dynamics Observatory (SDO, \citealt{pesnell2012solar}), \cite{panesar2021magnetic} found that most campfires are associated with cool plasma structures and magnetic flux cancellation events. They propose that flux cancellation triggers cool plasma eruptions, leading to the formation of campfires. A further study (\citealt{kahil2022magnetic}) using Polarimetric and Helioseismic Imager (PHI, \citealt{solanki2020polarimetric}) found that campfires are localized between bipolar magnetic features. They reported that these features show signs of magnetic flux cancellation, suggesting they might be driven by magnetic reconnection occurring at their footpoints. Recently, \cite{skan2023small} employed realistic magnetohydrodynamics simulations and forward synthesis of spectral lines (H$\alpha$ and \ion{Si}{4}) to investigate the formation of these small-scale loops. Their model suggests that loops within a bipolar system can generate numerous small-scale recurrent events heated to high temperatures. This occurs due to the rapid movement and rearrangement at the footpoints, enabling them to achieve coronal temperatures without requiring flux emergence. However, their study does not take into account all TR emissions.\\
Jets and jet-like eruptions are prevalent phenomena within the solar atmosphere, observable across both the solar disk and limb, and play a major role in the formation of transient loops. They occur across a wide range of spatial and temperature scales, from large-scale solar X-ray jets (\citealt{shibata1993observations}, \citealt{canfield1996halpha}, \citealt{shimojo1996statistical}) and EUV jets  (\citealt{nistico2009characteristics}, \citealt{panesar2016magnetic}, \citealt{sterling2016minifilament}) in the corona to smaller-scale jets (\citealt{shibata2007chromospheric}, \citealt{kuridze2011small}, \citealt{tian2014prevalence}, \citealt{patel2022hi}) in the lower solar atmosphere. In a recent study by \citet{sterling2015small}, they investigated many X-ray jets in polar coronal holes. They concluded that the formation of an X-ray jet is initiated by the destabilization of a sheared and twisted compact magnetic flux rope containing a minifilament situated at adjacent to minority-polarity side of a larger bipole. A disruption of the flux-rope field triggers an eruption, propelling the compact structure between the larger bipole and the ambient open field. As the eruption begins, internal reconnection occurs within the stretched-out legs of the minifilament field. This internal reconnection is responsible for the formation of flare arcade known as the ``Jet Bright Point" (JBP), which is observed at the edge of the base of the jet. The spire begins to take shape when the outer envelope of the erupting field, carrying the minifilament, initiates external reconnection with the open field situated on the far side of the large bipole. This external reconnection continues, injecting minifilament plasma along the open field and adding a new hot layer to the larger bipole. These processes of internal and external reconnection present a common mechanism for the formation of X-ray jets and loops at larger scales within the solar atmosphere. In a further study by \cite{sterling2016minifilament}, it was observed that AR coronal jets also exhibit evidence suggesting that they originate from small-scale eruptions, which are in turn prepared and triggered by magnetic flux cancellation.\\
In spite of several observational and simulation studies, we still lack a clear understanding of the formation mechanism of small-scale loops in the solar atmosphere. In this paper, we examined in detail two small-scale loops within an AR in the chromosphere and their TR and coronal signatures using data from Goode Solar Telescope (GST, \citealt{cao2010scientific}) at Big Bear Solar Observatory (BBSO), IRIS, and the Atmospheric Imaging Assembly (AIA, \citealt{lemen2012atmospheric}) and HMI onboard SDO and investigated their formation mechanism.
\section{Observations}
\begin{figure}[h!]
\renewcommand{\thefigure}{1}
\centering
\includegraphics[width=\textwidth]{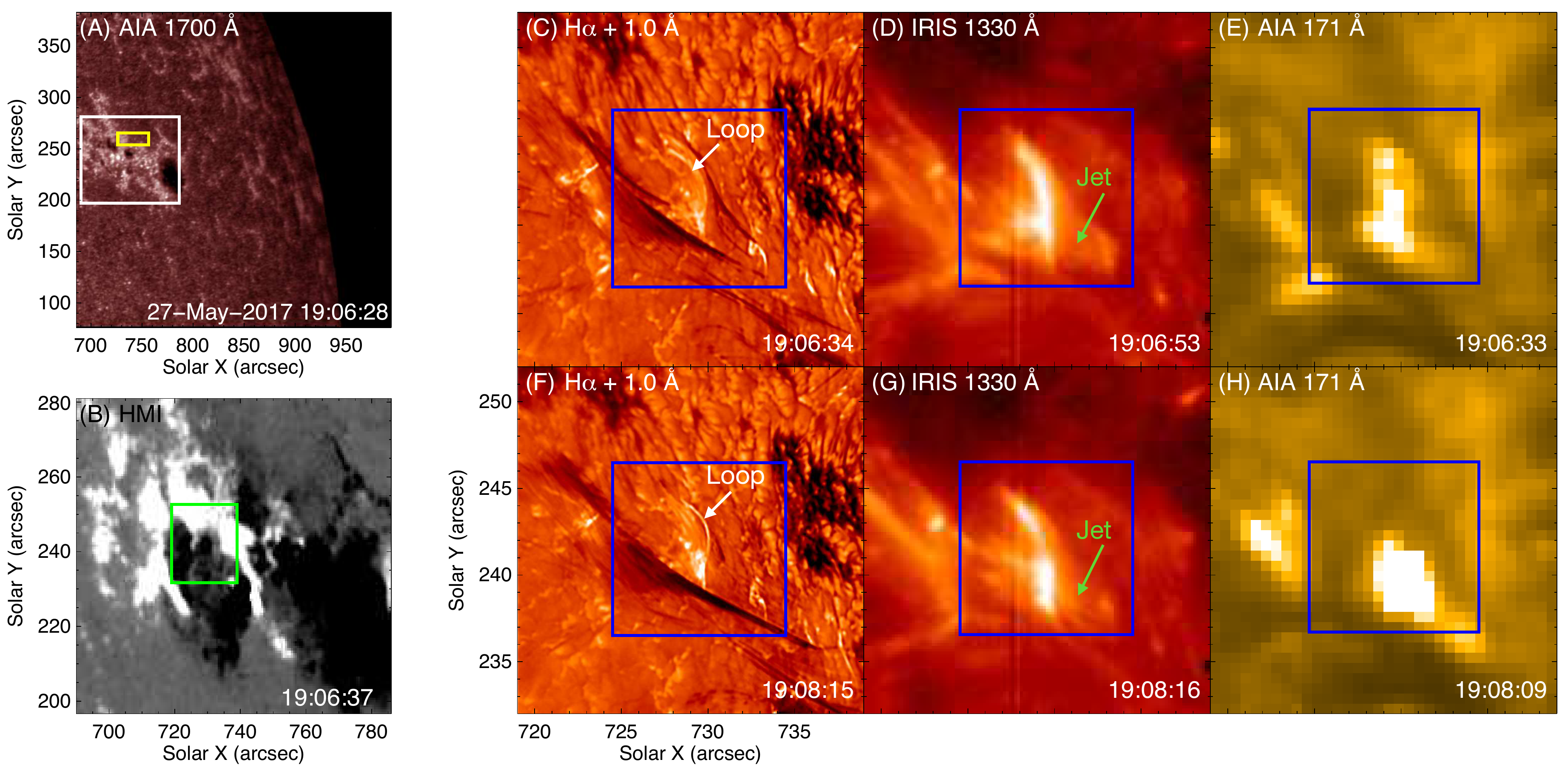}
\caption{(A): Shows an active region near the solar limb in AIA 1700~\AA\ at 19:06:28 UT on 2017 May 27. The yellow rectangle box marks the region of the plage where the reference profiles are taken in Figures~\ref{fig:spectra1},~\ref{fig:spectra2},~\ref{fig:demprofile} and \ref{fig:jet}. (B): Shows the HMI line-of-sight magnetic field map of the region marked by a white box in panel (A). The green box marks the FOV shown in panel (C)-(H). (C)-(E): Fine structure loop as observed by GST in red wing passband H$\alpha+$1.0~\AA, by IRIS in 1330~\AA\ and by AIA in 171~\AA\ at 19:06:34 UT. (F)-(H): Another fine structure loop as observed by GST in red wing passband H$\alpha+$1.0~\AA, by IRIS in 1330~\AA\ and by AIA in 171~\AA\ at 19:08:15 UT. The blue square box shows the FOV shown in Figure~\ref{fig:evolution}. North is to the top, south is toward the bottom, west is to the right, and east is to the left, in all solar figures in this paper. }
\label{fig: contextimage}
\end{figure}
We have analyzed the data obtained from the coordinated observations from the 1.6m GST at BBSO, IRIS, AIA and HMI on May 27, 2017. The Visible Imaging Spectrometer (VIS) of GST, basically a narrowband tunable Fabry–Pérot interferometer, captured images at H$\alpha$ core (0.0~\AA), and H$\alpha$ wings at $\pm$1, $\pm$0.8, $\pm$0.6, $\pm$0.4 and $\pm$0.2~\AA, sequentially, with a cadence of $\sim$ 53~s at each wavelength position for a duration 17:02-19:25 UT. The spatial resolution of the H$\alpha$ images is 0$''$.029 pix$^{-1}$. The emission in the H$\alpha$ core generally comes from the chromosphere, while in wings it mostly comes from lower atmospheric heights (\citealt{leenaarts2006dot}, \citeyear{2012}). For the IRIS data set, IRIS performed a large coarse 16-step raster with a $2''$ step size, covering a field-of-view (FOV) of $30''\times119''$ in NOAA AR 12659 for a duration 17:01-22:07 UT. The exposure time was 3.9~s for each slit position, and the step and raster cadence were 5.2~s and 83~s, respectively. The IRIS pointing coordinate for this observation was ($739''$, $246''$), close to the limb. We have used level 2 data for our analysis. IRIS provided Slit-Jaw Images (SJIs) at wavelengths of 2832~\AA, 2796~\AA, and 1330~\AA, with respective cadences of 83~s, 21~s, and 21~s. We have used SJIs taken in filters 2832~\AA\ and 1330~\AA. The IRIS 2832~\AA\ filter is mainly sensitive to the plasma emission from the upper photosphere at a temperature of $\sim$10$^{3.8}$~K, and the 1330~\AA\ filter is dominated by emission from the lower TR at $\sim$10$^{4.5}$~K temperature. The spatial resolution for SJIs and spectral images was $\sim$0$''$.166 pix$^{-1}$. The spectral dispersion was $\sim0.051$~\AA~pix$^{-1}$ in the near-ultraviolet band (NUV), whereas $\sim0.025$~\AA~pix$^{-1}$ in the far-ultraviolet band (FUV). The AIA onboard SDO  provided full disc images of the Sun in ten different filters with a spatial resolution of $\sim$0$''$.6 pix$^{-1}$. These images are obtained at a cadence of 12~s in Extreme Ultraviolet (EUV) filters and 24~s in the Ultraviolet (UV) filters. We used images taken in six EUV filters (94, 131, 171, 193, 211, and 335~\AA), which have centred wavelengths at various ionization stages of iron lines (\ion{Fe}{18} and \ion{Fe}{10}; \ion{Fe}{8}, \ion{Fe}{20} and \ion{Fe}{23}; \ion{Fe}{9}; \ion{Fe}{11}, \ion{Fe}{12} and \ion{Fe}{24}; \ion{Fe}{14}; and \ion{Fe}{16}) and also in one UV filter (1700~\AA), which samples the UV continuum emission around the temperature minimum region. These EUV wavelength channels, which are characterized by the formation temperature of their lines (log T[K]: 6.8; 5.6 and 7.0; 5.8; 6.2 and 7.3; 6.3; and 6.4), each have a broad 
\begin{figure}[h!]
\renewcommand{\thefigure}{2}
\centering
\includegraphics[width=\textwidth]{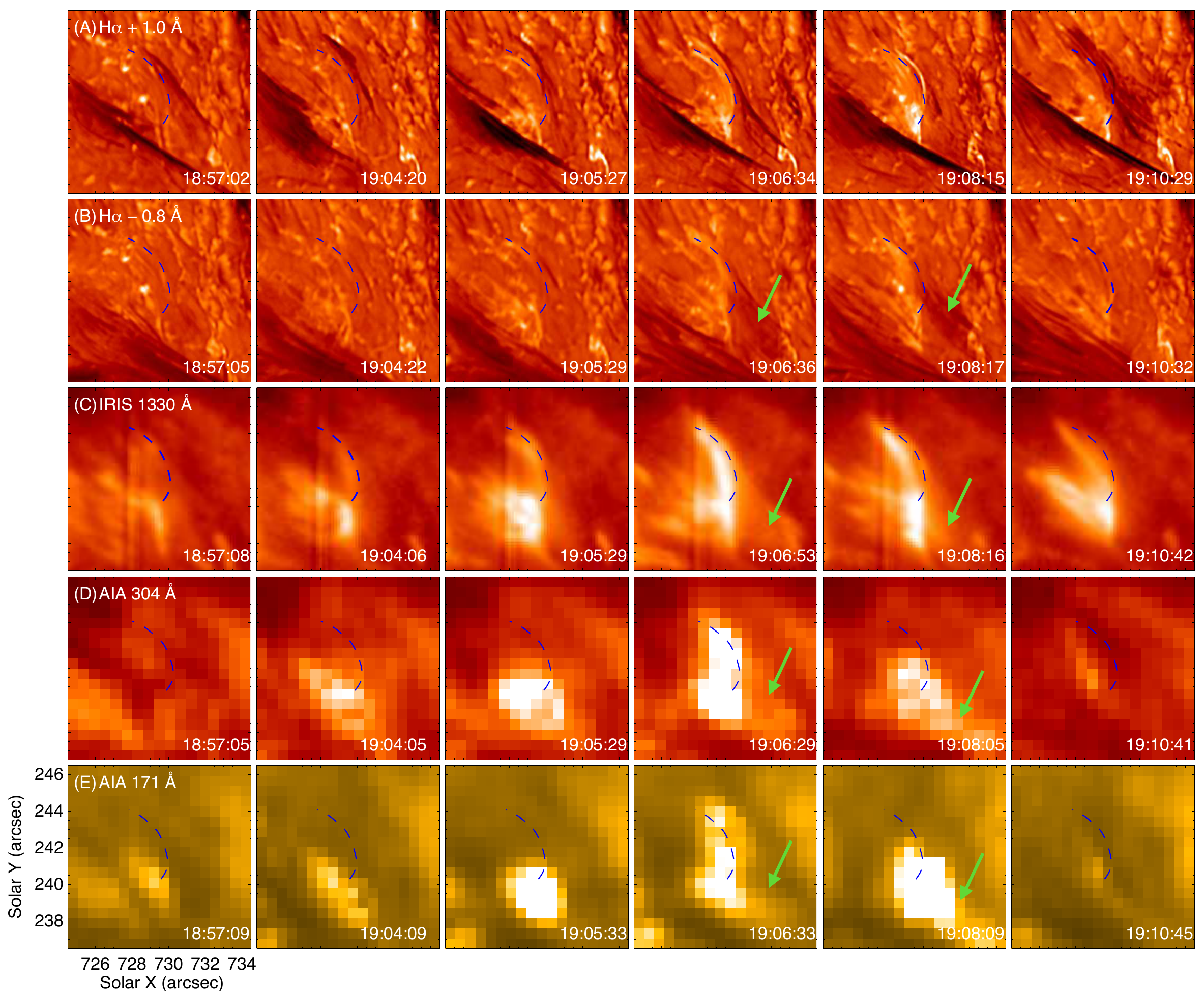}
\caption{Time evolution of loop. (A)-(E): Time evolution of loop in GST H$\alpha+$1.0~\AA, GST H$\alpha-$0.8~\AA, IRIS 1330~\AA, AIA 304~\AA\ and AIA 171~\AA, respectively. The blue dashed line marks the edge of the loop in GST H$\alpha+$1.0~\AA\ at 19:06:34 UT, which is plotted in all panels. The green arrow head is in the southwest side of the upward-erupting jet observed in GST H$\alpha-$0.8~\AA, IRIS 1330~\AA\ and across AIA passbands above the loops. An animation depicting the evolution of panels (A-E) from 18:50:19 UT to 19:14:25 UT provides an overview of their FOV.}
\label{fig:evolution}
\end{figure}
temperature response and together cover a wide range (0.1 to 20~MK) of plasma temperature (\citealt{lemen2012atmospheric}, \citealt{boerner2012initial}). HMI onboard SDO line-of-sight magnetograms of the solar photosphere with a spatial resolution of $\sim$0$''$.5 pix$^{-1}$ and a cadence of approximately 45~s. The AIA images were derotated to a reference time of 17:02:04 UT to eliminate the effects of solar rotation. Using a cross-correlation technique, the GST H$\alpha$ images captured at 11 wavelength positions were internally aligned. The GST H$\alpha$ images were carefully co-aligned with simultaneous AIA 1700~\AA\ images by matching specific observed features. Finally, the co-alignment of IRIS 1330~\AA\ SJIs with the GST dataset was achieved using 2832~\AA\ SJIs, comparing common photospheric features with H$\alpha$ $\pm$1~\AA\ wing images for each frame.
In our study, we examined the data for a duration of 17:02-19:25 UT. We selected a Region-of-Interest (ROI) spanning from $719''$ to $739''$ in solar X and from $232''$ to $252''$ in solar Y, which gives the green box panel B of Figure~\ref{fig: contextimage}, and is centered on the loops of interest.
\section{Data Analysis and Results} \label{DAR}
\noindent This study focuses on understanding the formation mechanism of small-scale loops in the lower solar atmosphere and the associated jet. We found two distinct small-scale loops located close to the limb in AR in our data where the IRIS slit was crossing. We studied these loops in detail using imaging and spectroscopy techniques and also examined their multi-wavelength behaviour. Figure~\ref{fig: contextimage} shows that these loops are visible in GST H$\alpha$~+~1.0~\AA\ and IRIS SJIs 1330~\AA\ at 19:06:34 UT and 19:08:15 UT, whereas the shape of these loops is much less resolved in AIA 304~\AA\ and 171~\AA\ observations.\\
We show the time evolution of these loops in Figure~\ref{fig:evolution}, with panels (A)-(E) depicting observations in GST H$\alpha$~+~1.0~\AA, GST~H$\alpha$~-~0.8~\AA, IRIS SJI 1330~\AA, AIA 304~\AA\ and 171~\AA, respectively. The blue dashed curve shows the edge of the
loop as observed in H$\alpha$ + 1.0~\AA\ at 19:06:34 UT, which is marked in all panels. The green arrow head marks the upward-erupting jet visible in H$\alpha$ - 0.8~\AA\ above the loop. No loop-like structure can be seen in 
\begin{figure}[h!]
\renewcommand{\thefigure}{3}
\centering
\includegraphics[width=\textwidth]{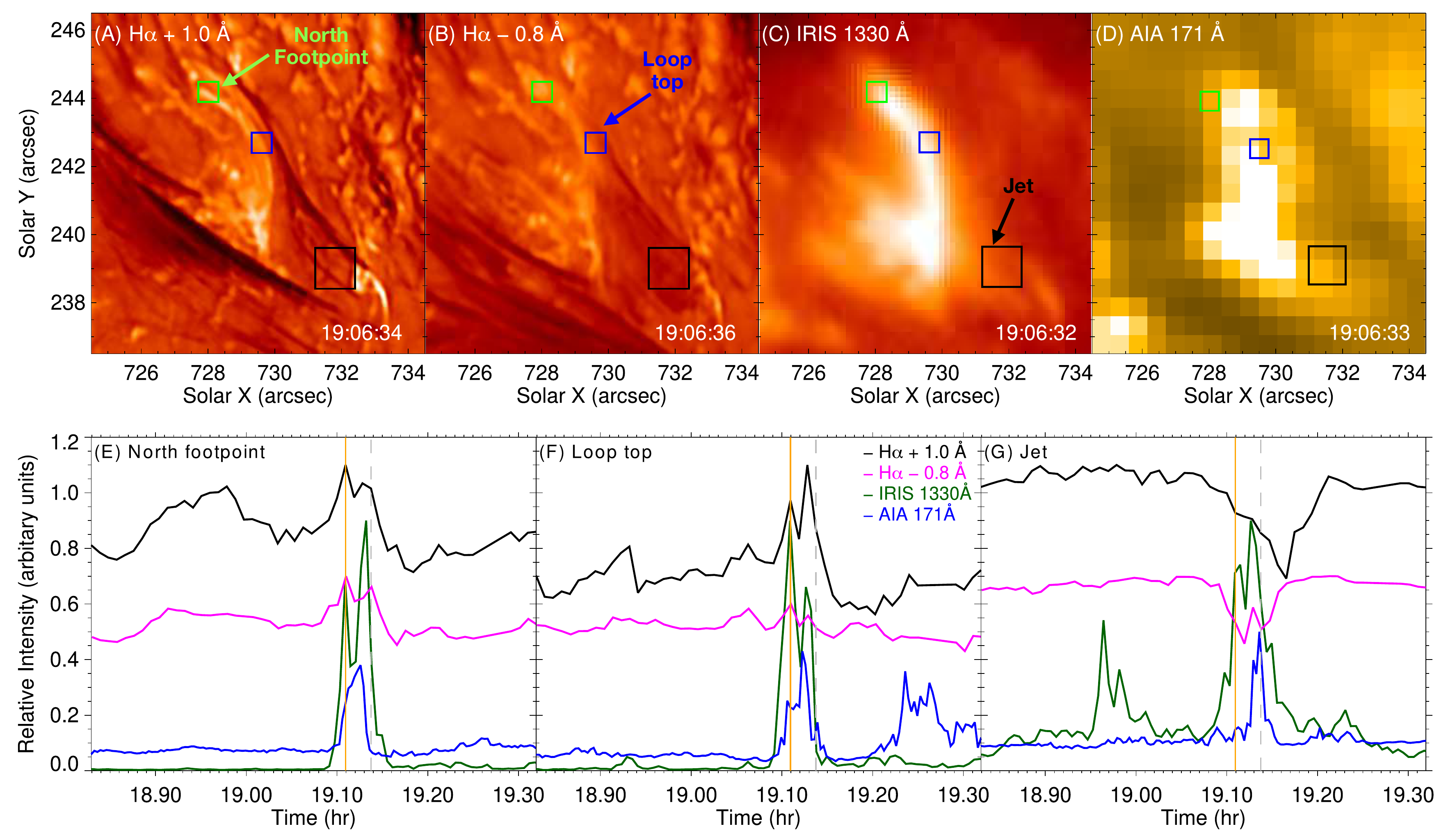}
\caption{Temporal evolution of
 H$\alpha$~+~1.0~\AA\ intensity,  H$\alpha$~-~0.8~\AA\ intensity, IRIS SJI 1330~\AA\ intensity and AIA 171~\AA\ intensity. The coloured curves indicate the different channels as labelled in panel (F). The green, blue and black boxes in panels (A)-(D)highlight the region used to calculate the north footpoint (panel E), loop top (panel F) and jet (panel G) intensity evolution, respectively. The yellow vertical line marks the appearance of the first loop at 19:06:34 UT, and the corresponding closest images in four channels are shown in panels (A)-(D). The grey dashed vertical line is placed at 19:08:15 UT. Intensities are in arbitrary units.}
\label{fig:synthesis}
\end{figure}
H$\alpha$ + 1.0~\AA\ at 19:04:20 UT. The loop appears from north around 19:05:27 UT in H$\alpha$ + 1.0~\AA\ and grows to about $\sim$3.6~Mm in length after more than a minute. We used the cubic spline interpolation method to determine the length and height of loops, assuming them to be semi-circular. First, we selected 5-6 data points along the loop, including the footpoints. Then, using the ``$spline.pro$'' function, we interpolated these data points to 100 points and fitted a cubic polynomial, ensuring that the curve is continuous and has continuous derivatives at each data point.  At 19:06:36 UT, a jet pointed to by a green arrow in panel B of Figure~\ref{fig:evolution} is visible in GST H$\alpha$~-~0.8~\AA, emanating above the loop. The south footpoint of the loop is much brighter than the rest of the loop, indicating stronger heating is happening in the lower solar atmosphere. At 19:08:15 UT, a slightly shifted, shorter loop appears whose length is $\sim$3.3~Mm. At 19:10:29 UT, the loop disappears from H$\alpha$ + 1.0~\AA. The SJI 1330~\AA\ evolution is slightly different from H$\alpha$ + 1.0~\AA. We noticed first the enhanced emission at the south footpoint becomes visible at 19:05:29 UT, and then the loop appears at 19:06:53 UT. There is an ejection at
\begin{figure}[h!]
\renewcommand{\thefigure}{4}
\centering
\includegraphics[width=\textwidth]{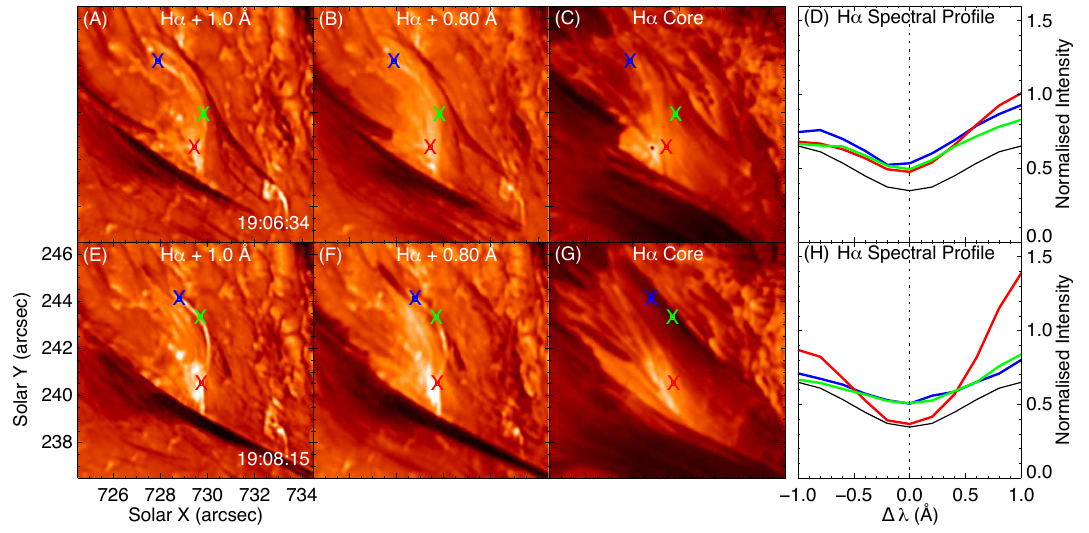}
\caption{(A)-(C): H$\alpha$ loop as observed by GST at 19:06:34 UT in H$\alpha$ red wing passbands H$\alpha$ + 1.0~\AA\ , H$\alpha$ + 0.8~\AA\ and H$\alpha$ core, respectively.(D): H$\alpha$ spectral profile at the location marked by red, green and blue cross (which marks the location of 3$\times$3 pixels square box) in panels (A)-(C). The black reference spectra are the spectra averaged over the entire FOV of GST. (E)-(H): Same profiles for another H$\alpha$ loop at 19:08:15 UT.}
\label{fig:halphaspectra}
\end{figure}
19:06:53 UT in an upward direction, which coincides with the jet seen in H$\alpha$~-~0.8~\AA\ at 19:06:36 UT. Another loop appears at 19:08:16 UT, which differs from the earlier one at 19:06:53 UT. We noticed that the second loop in IRIS SJI 1330~\AA\ at 19:08:16 UT does not coincide with the second loop observed in H$\alpha$ + 1.0~\AA\ at 19:08:15 UT, suggesting that these might be different structures. Therefore, we will call the loops observed in H$\alpha$ + 1.0~\AA\ as ``H$\alpha$ Loops" and those observed in IRIS SJI 1330~\AA\ as ``TR Loops". The evolution pattern seen in AIA 304~\AA\ and 171~\AA\ is similar to SJI 1330~\AA. However, the morphology of the brightening is unclear in AIA 304~\AA\ and 171~\AA\ due to the lower resolution of AIA. We observed the brightenings at the location of both the footpoints of these loops in AIA filters. The enhanced brightening disappears in AIA 304~\AA\ and 171~\AA\ around 19:10:41 UT and 19:10:45 UT, respectively. The loop shape of these fine structure loops becomes visible in H$\alpha$ + 1.0~\AA\ before it is clearly defined in SJI 1300, AIA 304 \AA\ and 171~\AA. The upward-erupting jet is also visible in AIA 304 \AA\ and 171 \AA\ filter. It is worth mentioning that there is significant activity prior to the formation of these loops, and we observed several brightenings at the location of the south footpoint in IRIS 1330 \AA, AIA 304 \AA\ and 171 \AA\ channels, which are likely to be associated with magnetic flux cancellation processes (\citealt{panesar2017magnetic}).\\
In Figure~\ref{fig:synthesis}, we illustrate the temporal evolution of intensities in the H$\alpha$~+~1.0~\AA,  H$\alpha$~-~0.8~\AA, IRIS SJI 1330~\AA, and AIA 171~\AA\ channels across different regions of interest within the loop and jet. The green, blue, and black boxes in panels (A)-(D) mark the regions used to track intensity changes at the north footpoint (panel E), loop top (panel F), and jet (panel G), respectively. The relative intensities within these regions in different channels are plotted with coloured curves in arbitrary units in panels (E)-(G), where black shows the intensity in H$\alpha$~+~1.0~\AA, magenta in H$\alpha$~-~0.8~\AA, green in IRIS SJI 130 \AA\ and blue in AIA 171 \AA\ channels. Around 19:06:34 UT (yellow solid line) and 19:08:15 UT (grey dashed line), the intensity profiles in the green and blue boxes show a sudden increase in all channels, indicating the appearance of loops. At the same time, the H$\alpha$~-~0.8~\AA\ intensity in the black box (magenta curve, panel G) shows a dip,  due to the jet erupting above the loop and appearing dark in the H$\alpha$~-~0.8~\AA\ blue wing. However, the intensity profiles within the black box for the SJI 1330 \AA\ and AIA 171 \AA\ channels show peaks attributed to the emission from the upward erupting jet. The
coherent peaks strongly indicate that the loops and the jet might be initiated from a common process. \\
For further analysis, we examined the H$\alpha$ spectral profile of these loops (panels D and H in Figure~\ref{fig:halphaspectra})
at three locations: the south footpoint (marked by a red cross), the north
footpoint (marked by a blue cross) and between the two footpoints (marked by a green cross). These crosses mark the
\begin{figure}[h!]
\centering
\includegraphics[width=\textwidth]{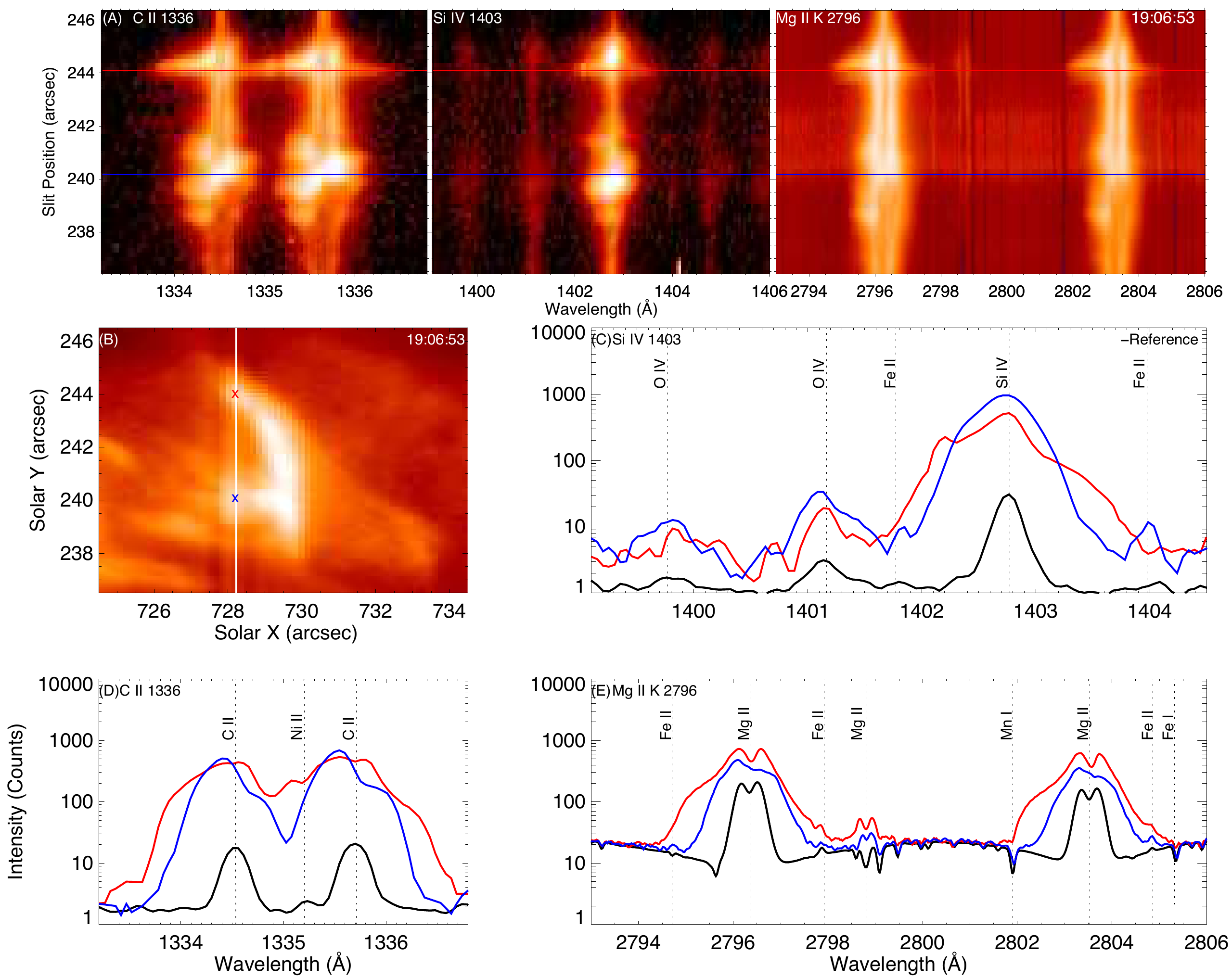}
\caption{IRIS spectra at the location of the loop at slit position 1. (A): Simultaneously taken IRIS spectral images in three spectral windows at slit position 1. (B): IRIS SJI 1330~\AA\ image taken at 19:06:53 UT. The white solid vertical line marks the IRIS slit position 1. (C)-(E): IRIS line profiles at the location of the loop marked by a red and blue cross in panel (B). The black line represents the reference line profile averaged over the plage region (yellow rectangle box in panel A of Figure~\ref{fig: contextimage}). The vertical dashed lines mark the reference wavelength of the elements taken from the atomic spectra database of NIST.}
\label{fig:spectra1}
\end{figure}
location of small square boxes (3$\times$3 pixels), and the average intensity within these square boxes is plotted in panels (D) and (H) of Figure~\ref{fig:halphaspectra} with respective colours. The black line in panels (D) and (H) is the reference profile obtained by taking the average intensity of the entire region for the whole observational duration. This reference profile was asymmetric in that the intensities of red wing is a little higher than in the blue wing. To remove this instrumental effect, we multiplied the intensities in the blue wing at each wavelength by a factor to make the H$\alpha$ spectral profile symmetric (\citealt{chen2019flame}). The profile at three marked crosses of the H$\alpha$ loop at 19:06:34 UT shows a large intensity enhancement at the wings, mainly the red wing. The lines are asymmetric at these positions. At 19:08:15 UT, the line profile at the south footpoint of the H$\alpha$ loop shows a large intensity enhancement in wings, mainly the red wing, and remains faint in the core. At the other two locations, the line
profile shows slightly higher intensity in the red wing compared to the blue, showing the lines are also asymmetric at this timestamp. Note that the jet above the H$\alpha$ loop at 19:08:15 UT is not visible in H$\alpha$ + 1.0~\AA\ and H$\alpha$ + 0.8~\AA\, but it is clearly visible in blue wing passband (panel B in Figure~\ref{fig:evolution}).
\begin{figure}[h!]
\centering
\includegraphics[width=\textwidth]{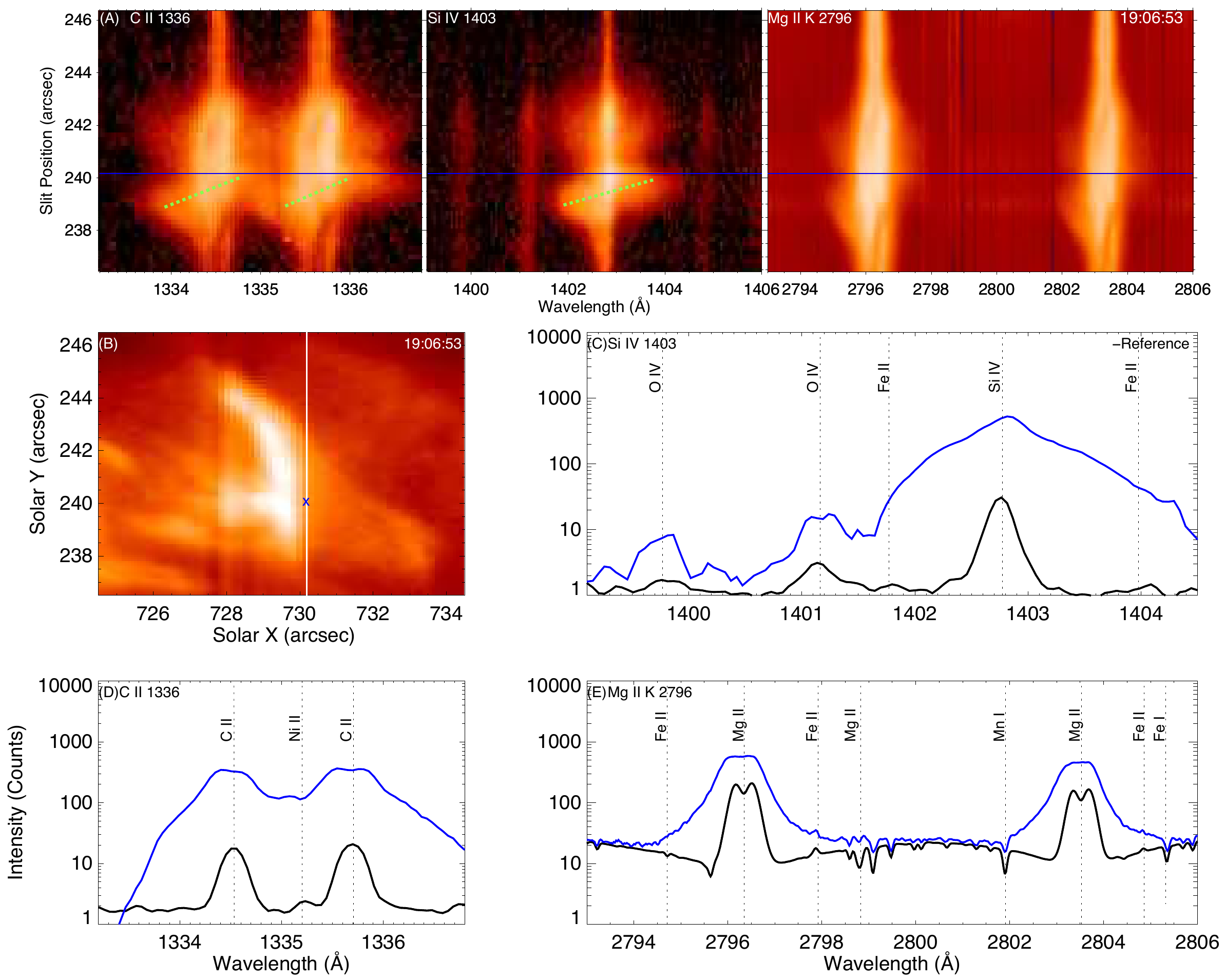}
\caption{IRIS spectra at the location of the loop at slit position 2. (A): Simultaneously taken IRIS spectral images in three spectral windows at slit position 2. (B): IRIS SJI 1330~\AA\ image taken at 19:06:53 UT. The white solid vertical line marks the IRIS slit position 2. (C)-(E): IRIS line profiles at the location of the loop marked by a blue cross in panel (B). The black line represents the reference line profile averaged over the plage region (yellow rectangle box in panel A of Figure~\ref{fig: contextimage}). The vertical dashed lines mark the reference wavelength of the elements taken from the atomic spectra database of NIST.}
\label{fig:spectra2}
\end{figure}
\\For further analysis, we also looked at the IRIS spectral profiles. IRIS performed a large coarse 16-step raster scan. However, during the lifetime of these loops, the slit crossed the location of their footpoints at two positions: at slit position 1, it crossed both footpoints and at slit position 2, it crossed slightly west of the south footpoint. The location of the IRIS slit at positions 1 and 2 is shown by a white solid vertical line in panel (B) of Figures~\ref{fig:spectra1} and~\ref{fig:spectra2}, respectively. We analyzed the IRIS spectral profile at the loop footpoints, marked by a red and blue cross in panel (B) of Figure~\ref{fig:spectra1}. The three spectral windows, namely \ion{Si}{4} 1403~\AA, \ion{C}{2} 1336~\AA\ and \ion{Mg}{2}~k 2796~\AA, at these locations are shown in panel (A).  As shown in Figure~\ref{fig:spectra1}, profiles of \ion{Si}{4} (panel C), \ion{C}{2} (panel D) and \ion{Mg}{2} (panel E) ions sampled at the location of footpoints of the loop at 19:06:53 UT are significantly broadened and greatly enhanced in both wings compared to the reference profiles. The reference profile (shown in black) was obtained by averaging the intensity of a plage region (marked by a yellow rectangle box in panel A of Figure~\ref{fig: contextimage}) for the entire observational period.\\ We noticed that at the location near the south footpoint of the loop during slit position 2, the profiles of \ion{Si}{4}, \ion{C}{2} and \ion{Mg}{2} display more broadening compared to line profiles at slit position 1, accompanied by a noticeable decrease in their peak intensities. We noticed the superposition of several chromospheric absorption lines on the greatly broadened wings of \ion{Si}{4} and \ion{C}{2} line profiles. These lines include \ion{Fe}{2} 1403.101~\AA, \ion{Fe}{2} 1403.255~\AA\ and  \ion{Ni}{2} 1335.203~\AA. These absorption lines are believed to result from some undisturbed upper chromosphere locations, indicating the likely presence of hotter gas beneath the upper chromosphere (\citealt{peter2014hot}). The IRIS data were initially smoothed by three pixels vertically along the slit to enhance the signal-to-noise ratio, due to which the chromospheric absorption features are less prominent in the spectra shown in Figures~\ref{fig:spectra1} and~\ref{fig:spectra2}. We noticed \ion{Fe}{2} 1403.977~\AA, 1401.771~\AA\ and 2794.711~\AA\ lines are present in emission near the location of the south footpoint of these loops at 19:06:53 UT, while these are absorption lines at the location of the north footpoint.
\begin{figure}[h!]
\renewcommand{\thefigure}{7}
\centering
\includegraphics[width=\textwidth]{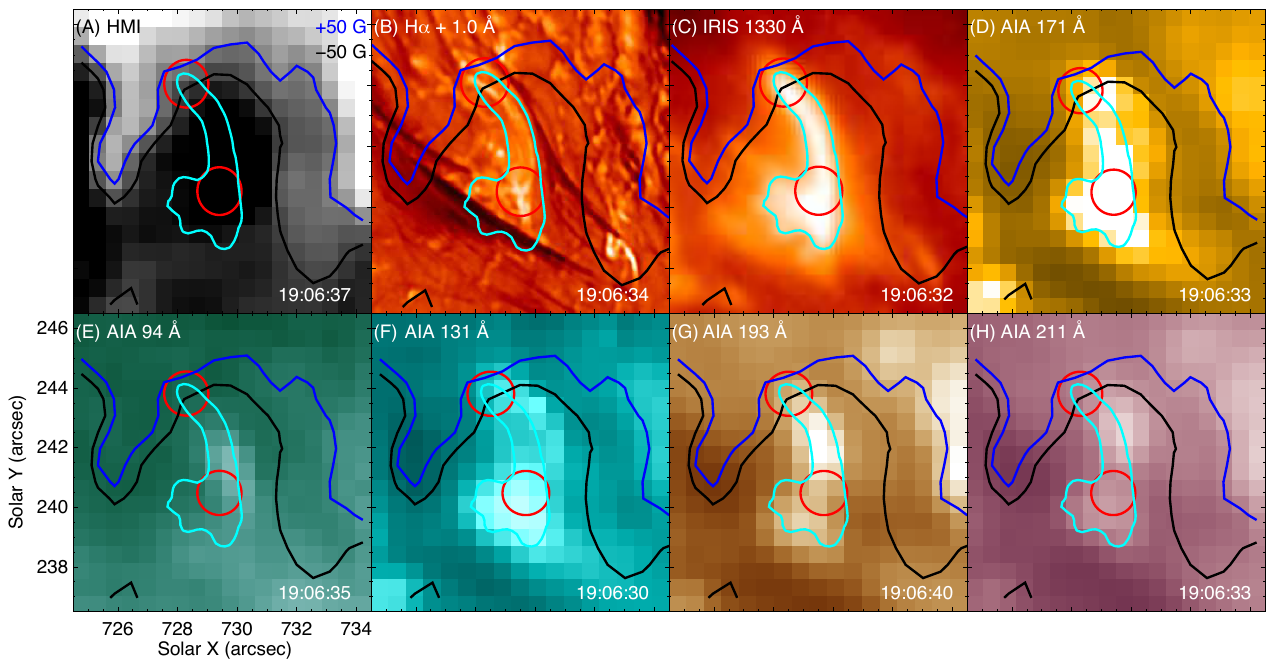}
\caption{ AIA, GST and IRIS observation of loop at 19:06:34 UT. (A): Shows the HMI line-of-sight magnetic field map of the region shown in previous figures. The black and blue contours correspond to the magnetic field strength of -50 G and +50 G, respectively. These contours are plotted in other panels also. (B): Shows H$\alpha$ loop in GST red wing passband H$\alpha$+1.0~\AA. (C): Shows TR loop in IRIS SJI 1330~\AA. The cyan contour marks the boundary of the TR loop as observed in IRIS SJI 1330~\AA\ at 19:06:32 UT, which is plotted in the other panels also. The red circle marks the footpoints of the H$\alpha$ loop in H$\alpha$+1.0~\AA\ at 19:06:34 UT. (D)-(H): Shows loop brightening in AIA EUV filters 171~\AA, 94~\AA, 131~\AA, 193~\AA\ and 211~\AA, respectively.  }
\label{fig:hmi}
\end{figure}
A notable feature is the presence of the \ion{Mg}{2} 2798.809~\AA\ emission line, which is a self-blend of two lines at 2798.754~\AA\ and 2798.822~\AA. Mostly, these lines appear as absorption lines, but they come into emission above the limb and in energetic phenomenon when strong heating occurs in the lower chromosphere (\citealt{pereira2015formation}).\\ We also noticed the self-absorption features present at the location of the north footpoint of the loop in the \ion{C}{2} and \ion{Mg}{2} lines at slit position 1 (panels D and E in Figure~\ref{fig:spectra1}). The self-absorption features in the \ion{C}{2} line are more prominent at the location of the north footpoint of the loop at 19:08:15 UT. This feature is not clear in \ion{Mg}{2} line at the south footpoint of the loop (shown by a blue line in panel (E) in Figure~\ref{fig:spectra1}) as the line core is much broadened and flattened. One of the possible explanations for this flattened line core could be that the line core is redshifted and giving rise to asymmetry, which makes the self-absorption feature unclear at the south footpoint. 
\\The coronal counterparts of these loops from AIA EUV images are shown in panels (D)-(H) of Figure~\ref{fig:hmi}, which show obvious emission at their location, mainly near the south footpoint. The plasma near the south footpoint of the H$\alpha$ loop, marked by a red circle, also shows emission in the AIA 94, 131, 193~\AA\ channels, which are sensitive to higher temperatures. These three channels exhibit a bimodal thermal response, which complicates the task of distinguishing the contribution of hot components from that of cooler structures along the line of sight. We conduct a Differential Emission Measure (DEM) analysis of EUV images of the loops to gain insights into their thermal structure and evolution using AIA EUV (94, 131, 171, 193, 211, 335~\AA) passbands.
\begin{figure}[h!]
\renewcommand{\thefigure}{8}
\centering
\includegraphics[width=\textwidth]{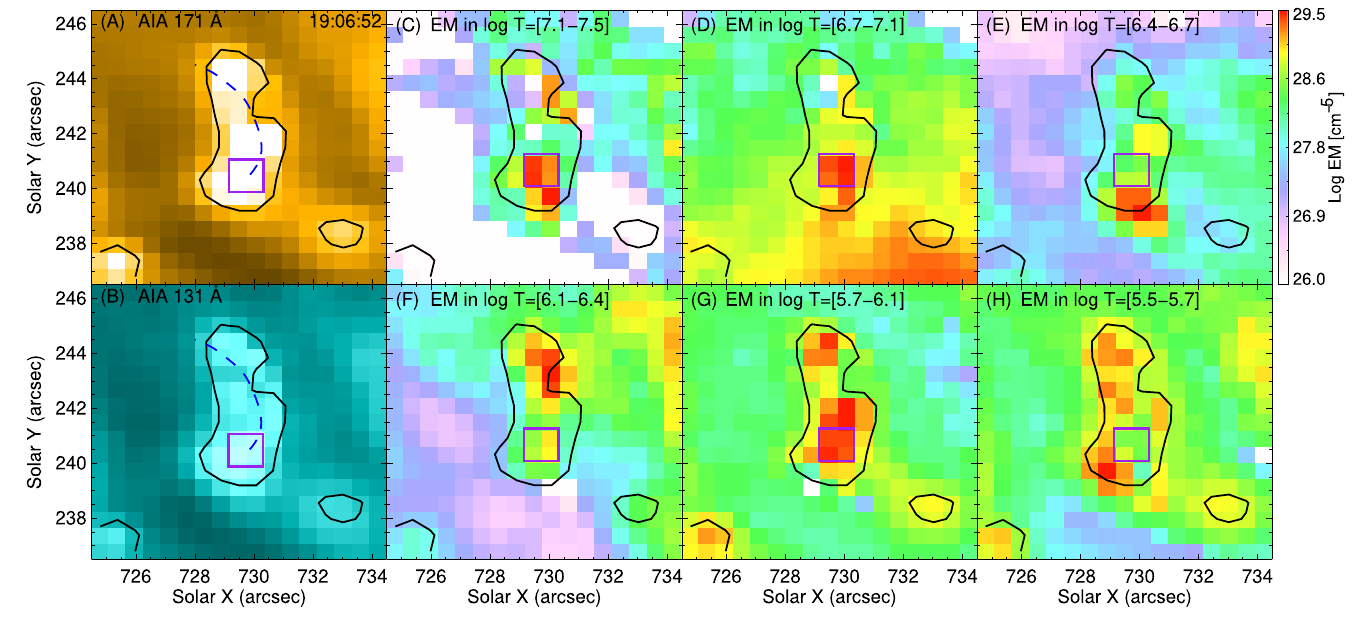}
\caption{Emission measure (EM) maps in different temperature ranges. (A)-(B): Observation at the location of the loop at 19:06:52 UT in AIA 171~\AA\ and 131~\AA\ filter, respectively. The black contour marks the loop in AIA 171~\AA\ at 19:06:52 UT. (C)-(H): Shows the EM maps in different temperature bins. The blue dashed line in panels (A)-(B) shows the edge of the H$\alpha$ loop in the H$\alpha$+1.0\AA\ passband at 19:06:34 UT. The purple square box (3 $\times$ 3 pixels) marks the location of hot plasma in the vicinity of the loop. }
\label{fig:demsolution}
\end{figure}
\\The Emission Measure (EM) analysis offers insights into the distribution of plasma at different temperatures along the observed line of sight. To deduce the EM using six AIA EUV channels data, we employed the inversion technique of \cite{cheung2015thermal}. Our inversion process used a temperature bin grid ranging from log T/K~=~5.5 to 7.5 with a bin width of log T~=~0.1. The EM in each temperature bin was determined through the $aia_{-}sparse_{-}em_{-}init.pro$ function in the SolarSoftware (SSWIDL) package. The temperature response functions for AIA channels are produced using $aia_{-}get_{-}response.pro$ incorporating timedepend, evenorm flags. To estimate errors in AIA intensities, $aia_{-}bp_{-}estimate_{-}error.pro$ is employed, accounting for various instrumental effects. The inversion procedure takes AIA count rates as input and yields the EM in each temperature bin. The resulting EM distributions are shown in Figure~\ref{fig:demsolution}. The EM maps in the vicinity of these loops show a clear signature of an enhanced EM near the south footpoint of the loop at bin temperatures log T/K~=~[7.1-7.5], [6.7-7.1], [5.7-6.1]. This location is marked as ``high-temperature brightening'' in panel (C) of Figure~\ref{fig:demsolution}. At the location of this footpoint, we also observed strong IRIS line broadening and a large intensity enhancement in the H$\alpha$ loop in H$\alpha$ + 1.0~\AA. To highlight this result from the spatial maps, we check DEM curves (Figure~\ref{fig:demprofile}) as a function of temperature for the region shown by the purple square box in Figure~\ref{fig:demsolution}. In the areas of the south footpoint of these loops, we see clear peaks at approximately log T/K~=~5.7, 6.5, 7.0, 7.1, which are marked with black arrows in Figure~\ref{fig:demprofile}. We also observed an enhanced EM at the location of the north footpoint of these loops at temperatures log T/K =[5.7-6.1], [6.1-6.4]. This location is marked as ``low-temperature brightening" in panel (G) of Figure~\ref{fig:demsolution}. We do not observe the jets in the DEM maps, which could be either because the temperature of the jet falls outside the range of log T=5.5-7.5, or due to insufficient EMs. At the location of the jet, the DEM analysis did not yield EMs in the higher temperature bins (panel (c), Figure~\ref{fig:demsolution}), which makes it difficult to determine the temperature of the jet. 
\begin{figure}[h!]
\renewcommand{\thefigure}{9}
\centering
\includegraphics[width=\textwidth]{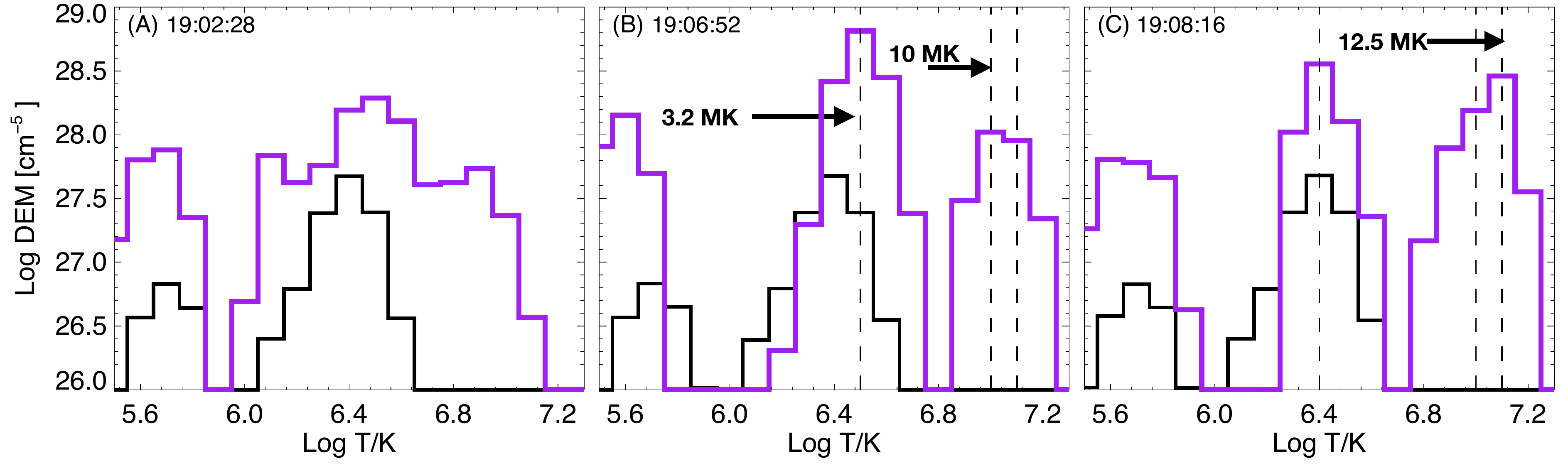}
\caption{Temporal evolution of DEM profile. (A)-(C): The purple curve shows the DEM profile at the location of the purple square box in Figure~\ref{fig:demsolution} averaged over 3$\times$3 pixels at 19:02:28 UT, 19:06:52 UT and 19:08:16 UT, respectively. The black curve is the DEM profile averaged over the plage region (yellow rectangle box in panel A of Figure~\ref{fig: contextimage}). }
\label{fig:demprofile}
\end{figure}
\section{Discussion} \label{Discussion}
\begin{figure}[h!] 
\renewcommand{\thefigure}{10}
\centering
\includegraphics[width=\textwidth]{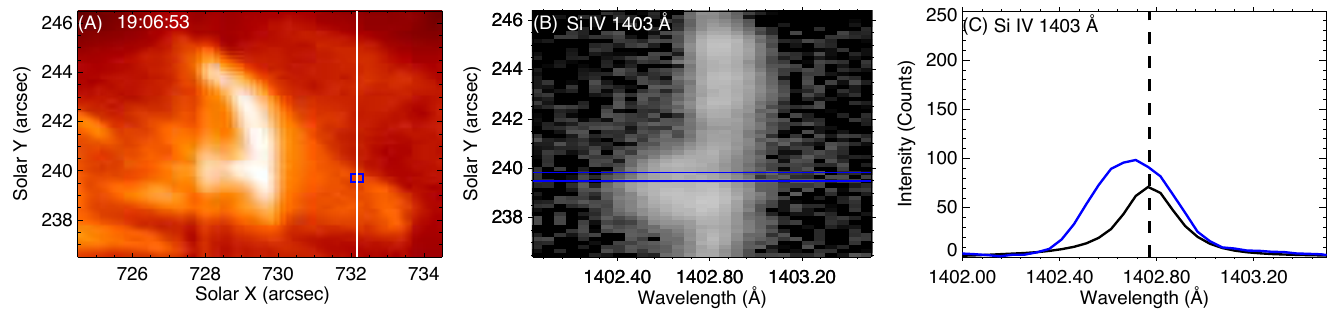}
\caption{IRIS spectra of jet at slit position 3.(A): IRIS SJI 1330 \AA\ image taken at 19:06:53 UT. The solid vertical white line marks
the IRIS slit position 3, which crosses the jet. (B): Shows the Si IV 1403 \AA\ spectral window at slit position 3. (C): IRIS line profile of jet at the location marked by blue box in panel (A). The black line profile is the reference line profile averaged over the plage region (yellow rectangle box in panel A of Figure~\ref{fig: contextimage}).}
\label{fig:jet}
\end{figure}

\begin{figure}[h!]
\renewcommand{\thefigure}{12}
\centering
\includegraphics[width=\textwidth]{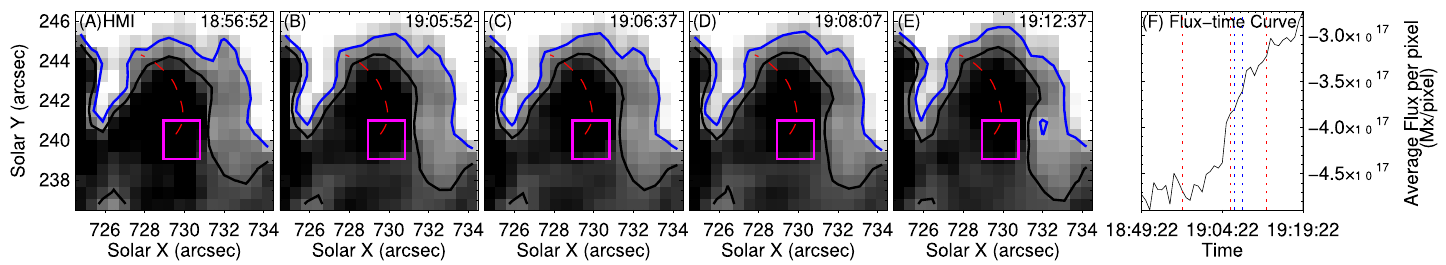}
\caption{Decrease of negative magnetic flux near the loop's south footpoint in HMI magnetograms before and during the loops and jet. The blue and black contours correspond to the magnetic strength of +50 and -50~G, respectively. The red curve marks the edge of the H$\alpha$ loop in GST H$\alpha$ + 1.0~\AA\ at 19:06:34 UT. The magenta square box highlights the region around the south footpoint of these loops within which the total magnetic flux is taken, and the resulting flux profile for this region is shown in Panel (F). The vertical dashed lines in panel (F) mark the time shown in panel (A)-(E). The vertically dashed blue lines mark the time of two loops.   }
\label{fig:hmievolution}
\end{figure}

Using coordinated observations of GST, IRIS, AIA and HMI, we studied the formation mechanism of small-scale loops located in an AR. High-resolution H$\alpha$ observations reveal that these H$\alpha$ loops have full lengths of $\sim$3.5~Mm. We also find a maximum height of $\sim$1~Mm, indicating that these loops are located within the chromosphere region. Our analysis showed that the H$\alpha$ loops primarily appear as bright emission structures in the far red wing of H$\alpha$, and they do not show up prominently in H$\alpha$ blue wing and core images, causing asymmetries in the H$\alpha$ line. This suggests that plasma in H$\alpha$ loops is redshifted and slightly hotter than ambient chromospheric plasma. \cite{pereira2018chromospheric} studied low-lying loops in the quiet Sun that were visible solely in the far blue or red wings of H$\alpha$ but appeared as absorption features. Our observations show brightening at the south footpoint with a marked enhancement in the H$\alpha$ spectral profile predominantly in the red wing (panels D and H in Figure~\ref{fig:halphaspectra}). However, \cite{pereira2018chromospheric} did not observe any brightenings or evidence of heating at the footpoint.\\
Detailed spectroscopic analysis at the location of these loops using IRIS revealed that throughout the loops lifetime, the spectral profiles of \ion{C}{2}, \ion{Si}{4} and \ion{Mg}{2} similarly display significant broadening beyond 200 km s$^{-1}$ in both wings, indicating the presence of reconnection flows within a small region (\citealt{dere1989explosive}, \citealt{innes1997bi}, \citealt{chae1998photospheric}). \cite{innes2015iris} studied the broadening of the IRIS \ion{Si}{4} line, and they proposed that the increased emission in the line wings, along with strong enhancement of the line cores, is due to the small-scale fast magnetic reconnection proceeding via plasmoid instability along the current sheet. Ellerman Bombs and UV bursts also show similar IRIS profiles (\citealt{peter2014hot}, \citealt{chen2019flame}, \citealt{ortiz2020ellerman}). However, in these instances, the spectral profiles exhibit prominent chromospheric absorption features as well as self-absorption features. In contrast, at the location of our loops studied here, these features are not as prominent, which might be due to intense magnetic reconnection in the lower chromosphere occurring near the footpoint (\citealt{pereira2015formation}), which heats nearly all the chromospheric material situated above it. \cite{peter2019plasmoid} provided a plasmoid-mediated reconnection model in the UV bursts and explosive events, which shows similar spectral profiles to these loops. The tilted spectra (\ion{C}{2} and \ion{Si}{4}) near the location of the south footpoint of loops (indicated by a green dashed line in panel A in Figure~\ref{fig:spectra2}) reveal a twist or bi-directional motion in them (\citealt{joshi2021multi}). The upward-erupting jets in GST blue wing passbands (panel B in Figure~\ref{fig:evolution}) above the H$\alpha$ loops are also observed in IRIS SJI 1330~\AA\ and AIA EUV filters (panels C-H in Figure~\ref{fig:hmi}), and the corresponding blue-wing enhancement of the IRIS \ion{Si}{4} spectral profile at slit position 3 (Figure~\ref{fig:jet}), which is at the location of jet, confirms the jet's upward flow.
\begin{figure}[h!] 
\renewcommand{\thefigure}{11}
\centering
\includegraphics[width=\textwidth]{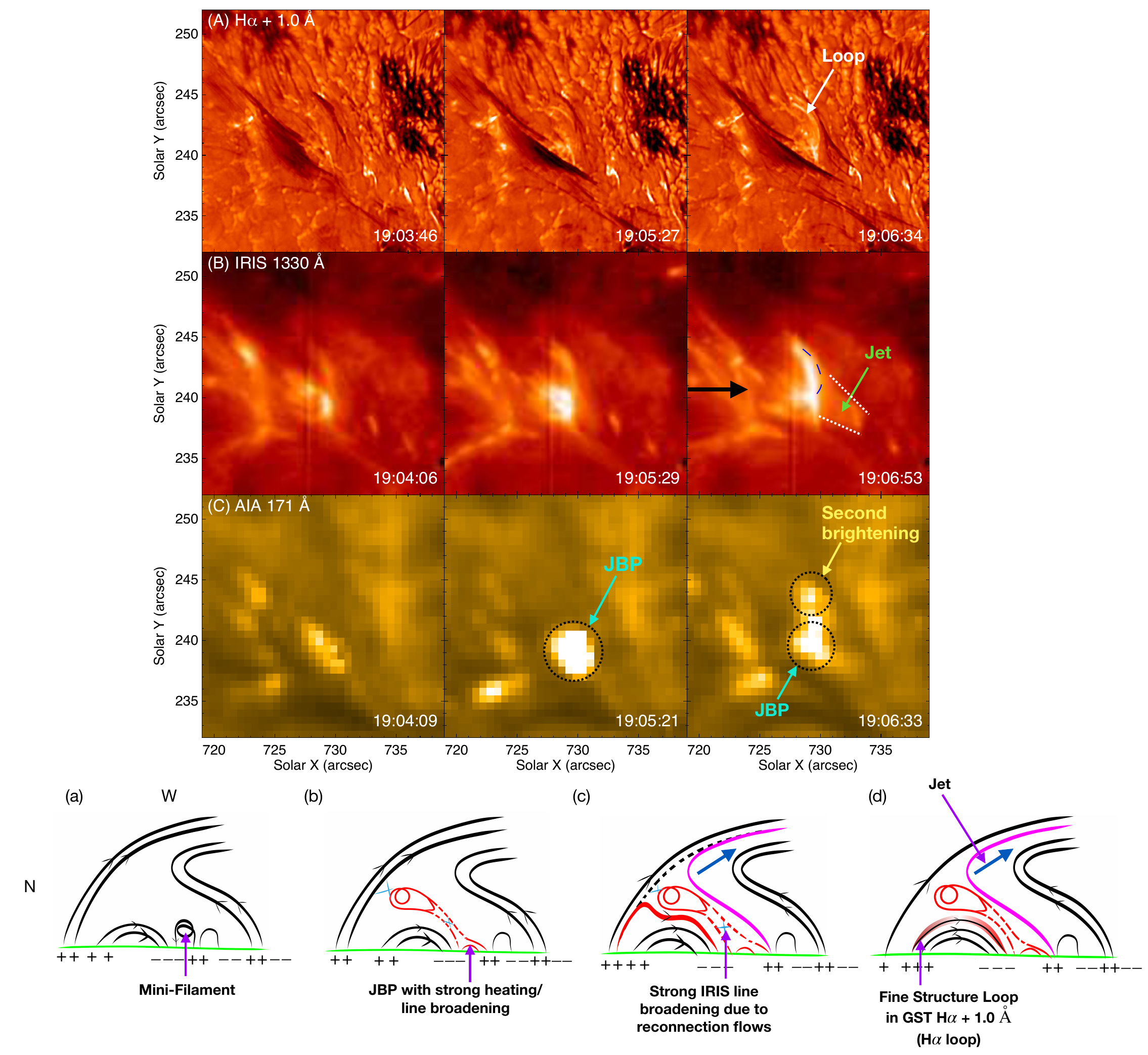}
\caption{An illustration of a potential mechanism for the formation of loops and the associated jet in 2D, based on the minifilament eruption model. Black lines show the magnetic field, with arrow heads indicating their polarities. Reconnection locations are marked by a light blue cross. The green line is the photosphere. (a) shows the scenario around 19:04:06 UT. (b) shows the scenario around 19:05:29 UT. (c) shows the stage of fastest internal and external reconnection. (d) shows the scenario around 19:06:53 UT. GST H$\alpha$ + 1.0~\AA, IRIS SJI 1330~\AA, AIA 171~\AA\ observations with closest timestamps are shown in panel (A)-(C). The black arrow in panel (B) marks the viewing direction for the schematic representations in panels (a)-(d). The positive field lines at both extreme ends represent the anticipated coronal magnetic connections to the nearby negative flux.}
\label{fig:cartoon}
\end{figure}
\\Using DEM analysis (\citealt{cheung2015thermal}), we found an enhanced EM in the proximity of the south footpoint of these loops and from the DEM profile, with temperature peaks at log T/K~=~5.7, 6.5, 7.0, 7.1 in the vicinity of these loops. This region coincides with the strong broadening due to reconnection flows in IRIS spectral profiles at the location of the TR loop in SJI 1330 \AA. This indicates that this might be the region of magnetic reconnection and plasmoid formation, due to which the plasma was heated to coronal temperatures at this location. However, a recent study by \cite{athiray2024can} showed that EM distributions derived solely from AIA data might overestimate the amount of high-temperature (log T $>$ 6.4) plasma in the solar corona. It has been suggested in previous studies (\citealt{hanneman2014thermal}, \citealt{cheung2015thermal}) that EM distribution incorporating X-Ray Telescope (XRT, \citealt{golub2008x}) data provide more accurate results compared to those without XRT data.\\
Based on our findings, we deduce a simple schematic representation depicted in Figure~\ref{fig:cartoon} for the formation of the small-scale loops and the generation of the jet, in accordance with the minifilament eruption model in 2D (\citealt{sterling2015small}). The average length of the minifilaments in \cite{sterling2015small} study was $\sim$8~Mm. However, recent studies (\citealt{sterling2016microfilament}, \citealt{sterling2020possible}) postulated that even smaller-scale filaments that they call ``microfilaments'' in postulated spicule-sized cases could also drive small-scale jets. We consider the scenario at the beginning (panel a in Figure~\ref{fig:cartoon}) as two sets of neighbouring magnetic bipole that exist next to each other. The larger bipole stands alongside a smaller bipole that has tightly twisted magnetic field containing a minifilament. When the field containing the minifilament becomes unstable due to some process, it erupts outward, guided between the larger bipole and the surrounding far-reaching magnetic field. As the minifilament rises, reconnection occurs below it within its stretched magnetic field legs, known as ``internal reconnection" (panel b in Figure~\ref{fig:cartoon}), forming a JBP. This is consistent with the initial brightening we observed near the location of the south footpoint of the loop around 19:05:21 UT in the time evolution (panel C in Figure~\ref{fig:cartoon}) of AIA 171~\AA. At this location, we also observed strong IRIS line broadening due to reconnection flows (Figures~\ref{fig:spectra1} and~\ref{fig:spectra2}) consistent with internal magnetic reconnection. Although we did not observe a minifilament, presumably due to the limitation of our observations, which includes temperature coverage of H$\alpha$ bandpass filter, narrow-band H$\alpha$ filtergram (0.07~\AA) and cadence ($\sim$53~s), the tilted spectra (marked by green dashed lines in panel A of Figure~\ref{fig:spectra2}) near the location of the south footpoint of the loop nonetheless suggests the possibility of the existence of the twisted minifilament flux rope, guiding between the larger bipole and the far-reaching magnetic field line. The corresponding observational images at this stage are shown in the middle column of Figure~\ref{fig:cartoon}. As the outer envelope of the erupting minifilament encounters the external field on the opposite side of the larger bipole, it undergoes an ``external reconnection". This external reconnection produces a jet along the far-reaching field line and also adds a heated layer to the larger bipole (red loop in panel c of Figure~\ref{fig:cartoon}). Panel d shows the relaxing phase of the loop and jet. The redshifted H$\alpha$ loop observed in  H$\alpha$ + 1.0~\AA\ at 19:06:34 UT (shown in Figure~\ref{fig:halphaspectra}) corresponds to this newly formed hot loop after it cools to chromospheric temperature. When the minifilament erupts, it undergoes multiple reconnections, and the second loop at 19:08:15 UT is likely due to the subsequent external reconnection. We also observed the counterparts of H$\alpha$ loop in IRIS SJIs (panel C in Figure~\ref{fig:evolution}), but there is a slight morphological difference, which could be due to the additional plasma dynamics in TR due to the plasmoid-mediated reconnection above the newly formed H$\alpha$ loop. Additionally, in the AIA 171~\AA\ filter, we observe a second brightening at the location of the north footpoint of the loop. The jet produced along the external far-reaching field line from the external reconnection is visible in H$\alpha$ blue wing passbands (indicated by a green arrow in panel B of Figure~\ref{fig:evolution}). This jet is also observed in IRIS SJI 1330~\AA\ and all AIA EUV filters. The corresponding observed images at this stage (panel c in Figure~\ref{fig:cartoon}) are shown in the right column of Figure~\ref{fig:cartoon}. 
\\All observational evidences we discussed are consistent with the scenario illustrated in the cartoon (Figure~\ref{fig:cartoon}). 
Although we lack high-resolution magnetic field measurements, we examined available magnetogram data to infer the presence of cancelling negative magnetic flux. The structures in our study are close to the limb, which restricts our ability to inspect their magnetic environment. The HMI magnetogram (panel A in Figure~\ref{fig:hmi}) shows that the south footpoint of the loop is dominated by negative polarity flux, and the north footpoint lies at the boundary of opposite (positive) polarity flux. The lower resolution of the magnetic field measurements might have affected the detection of weak magnetic elements with positive polarity near the southern footpoint of the loops. However, at approximately 18:56:52 UT, a distinctive streak of decreased negative flux becomes evident at the south footpoint of these loops. In Figure~\ref{fig:hmievolution}, we show the HMI magnetogram at 
five timestamps and a corresponding magnetic flux time profile for the region enclosed by the magenta square box around the south footpoint. There is a fast decrease in negative flux following the 18:56:52 UT timestamp (panel F in Figure~\ref{fig:hmievolution}), consistent with flux cancellation. This indicates the presence of weak positive flux embedded within the surrounding big negative region on the left, near the base of the jet. We also checked flux-time variation using three different box sizes centered at [729$''$,~239$''$], [729.5$''$,~239.5$''$], and [730$''$,~240$''$] with dimensions of 4$''\times$4$''$, 3$''\times$3$''$, 2$''\times$2$''$, respectively. We found that the overall trend of decreasing negative flux remains consistent. Recently \cite{nobrega2024small} highlighted the importance of a resolution at least 4 to 5 times larger than that of HMI to properly resolve small-scale magnetic flux emergence episodes; our findings here suggest that similarly high resolution is also needed to resolve fine-scale flux cancellation. Our findings are consistent with the minifilament eruption as the driving force behind the formation of the heated loops and the associated jet.
\section{Conclusions}
Using high-resolution ground and space-based data, we investigated the formation and dynamics of small-scale loops and analyzed their multi-wavelength behaviour. We analyzed such loops in the chromosphere and their TR and corona counterparts using both imaging and spectroscopy techniques for the first time. We found that the loops appear as bright emission structures in the far red wing of H$\alpha$. The IRIS spectral profiles (\ion{C}{2}, \ion{Si}{4}, \ion{Mg}{4}) at the location of the loops show strong broadening in both wings and the intensity of the emission lines closely resembles that observed in local heating events like UV bursts (\citealt{young2018solar}). This similarity suggests that the observed broadening could arise from small-scale magnetic reconnection processes, possibly via plasmoid instability along the current sheet (\citealt{innes2015iris}). Our DEM analysis reveals that the plasma near the footpoint of these loops reaches temperatures of a million degrees. This heating contributes to the brightening observed in AIA EUV filters. Based on our observations, we found that these small-scale loops form in accordance with the minifilament eruption model for coronal jets (\citealt{sterling2015small}). Therefore, we discovered a similarity of the mechanism driving the formation of small-scale loops and jets to that of larger-scale X-ray jets. Our study highlights the significance of using high-resolution magnetograms to identify small-scale magnetic flux cancellation events. Higher-resolution magnetograms are crucial for understanding and correlating them with subsequent eruptive phenomena. Without high-resolution magnetic field data, our understanding could be incomplete, possibly leading to misinterpretations.\\\\
Acknowledgements. We thank Mr.~Anand Parikh for his initial exploration of data in his master's thesis. We gratefully acknowledge the use of data from the Goode Solar Telescope (GST) of the Big Bear Solar Observatory (BBSO). BBSO operation is supported by the US NSF AGS 2309939 grant and the New
Jersey Institute of Technology. The GST operation is partly supported
by the Korea Astronomy and Space Science Institute and the Seoul
National University. IRIS is a NASA small explorer mission developed and operated by LMSAL with mission operations executed at NASA Ames Research Center and major contributions to downlink communications funded by ESA and the Norwegian Space Centre. A.C.S. and R.L.M. received funding from the Heliophysics Division of NASA’s Science Mission Directorate through the Heliophysics Supporting Research (HSR) Program. V.Y. acknowledges the support from US NSF AGS grants 2401229, 2408174, 2300341, NSF AST-2108235 grant, and NASA 80NSSC24K1914 grant.
\bibliography{FSL}{ }
\bibliographystyle{aasjournal}
\end{document}